# Enhancing Automotive Security with a Hybrid Approach towards Universal Intrusion Detection System

Md Rezanur Islam[a], Mahdi Sahlabadi[b], Keunkyoung Kim[a] and Kangbin Yim[b,*]

[a]*Dept. of Software Convergence, Soonchunhyang University, Soonchunhyang-ro 22, Asan-si, 31538, South Korea*
[b]*Dept. of Information Security Engineering, Soonchunhyang University, Soonchunhyang-ro 22, Asan-si, 31538, South Korea*

**ABSTRACT**

Security measures are essential in the automotive industry to detect intrusions in-vehicle networks. However, developing a one-size-fits-all Intrusion Detection System (IDS) is challenging because each vehicle has unique data profiles. This is due to the complex and dynamic nature of the data generated by vehicles regarding their model, driving style, test environment, and firmware update. To address this issue, a universal IDS has been developed that can be applied to all types of vehicles without the need for customization. Unlike conventional IDSs, the universal IDS can adapt to evolving data security issues resulting from firmware updates. In this study, a new hybrid approach has been developed, combining Pearson correlation with deep learning techniques. This approach has been tested using data obtained from four distinct mechanical and electronic vehicles, including Tesla, Sonata, and two Kia models. The data has been combined into two frequency datasets, and wavelet transformation has been employed to convert them into the frequency domain, enhancing generalizability. Additionally, a statistical method based on independent rule-based systems using Pearson correlation has been utilized to improve system performance. The system has been compared with eight different IDSs, three of which utilize the universal approach, while the remaining five are based on conventional techniques. The accuracy of each system has been evaluated through benchmarking, and the results demonstrate that the hybrid system effectively detects intrusions in various vehicle models.

## 1. Introduction

Artificial intelligence has brought about a revolution in the automotive industry, allowing for the development of fully automated vehicles Contreras-Castillo, Zeadally and Guerrero-Ibañez (2017). However, as technology continues to advance, there is a growing concern regarding the security of In-Vehicle Networks (IVNs) Yu, Liu, Xie, Li, Liu and Yang (2022); Kim, Oh, Yim, Sahlabadi and Shukur (2023). Industry is obligated to recognize the risks associated with IVNs and take necessary precautions to ensure the safety of both drivers and passengers. Modern vehicles are equipped with integrated electronic systems and sensors to enrich the connectivity feature. This phenomenon extends the surface of the attack and brings attention to the Intrusion Detection System (IDS). However, because of the nature of the data generated by vehicles, it is difficult to build an IDS solution that is appropriate for every scenario Hoang, Islam, Yim and Kim (2023) for two reasons: First, every vehicle has a unique network architecture that generates very varied data profiles Choi, Lee, Joo, Jo and Lee (2021) for a scenario; second, the data generations vary depending on the circumstances and driving styles Gazdag, Lestyán, Remeli, Ács, Holczer and Biczók (2023). It is challenging to establish an extensive dataset that covers a wide range of scenarios because of these factors. There are few available datasets that present very limited scenarios. The researchers primary motivation for avoiding cross-validation is their awareness of that. The majority of the AI algorithms appear to operate with high-accuracy matrixes because the dataset contains relatively little situational data, yet the model is not generalizable. It is critical to use an IDS that can handle these variations and provide accurate results.

IDSs primarily monitor Controller Area Network (CAN) data because it is the main communication protocol within IVN Jeong, Kim, Han and Kwak (2023). IDSs face challenges when deployed across different CAN datasets Liu, Zhang, Sun and Shi (2017). CAN networks vary in function and design to meet specific vehicle needs and cater to the complexity of electronic and autonomous vehicles. For instance, Kia has three distinct CAN channels with Electronic Control Unit (ECU) numbers of approximately 100, each responsible for different functions An, Park, Oh, Kim and Yim (2021). CAN networks handle crucial functions with B-CAN, C-CAN, and M-CAN to manage body, chassis, and multimedia systems, respectively. These networks vary in speed and priority to cater to specific vehicle needs. Although the Kia models share the same channel, their internal physical CAN ID and payload differ Hoang et al. (2023). Similarly, the BMW K-CAN, V-CAN, and K2-CAN consist of different components according to the manufacturer's internal functional design Cai, Wang, Zhang, Gruffke and Schweppe (2019). A significant difference between mechanical and electronic vehicles is their IVN architectures. This is because manufacturers offer various features to attract customers, and autonomous vehicles require many sensors to ensure safety Liu, Lu, Zhong, Wu, Yao, Zhang and Shi (2020). For example, Tesla's ECUs vary in number, with each version around four ECUs Vdovic, Babic and Podobnik (2019), and it has nine types of CAN channels Tesla Owners

---

*Corresponding author.
yim@sch.ac.kr (K. Yim)
ORCID(s): 0000-0002-1183-7741 (M.R. Islam); 0000-0002-1361-1455 (K. Yim)

Online Forum (2016). CAN networks undoubtedly vary, as demonstrated by these industrial examples.

Despite the importance of improving vehicle safety, the deployment of conventional IDS has faced significant challenges due to variations in network architecture and data generation practices among manufacturers. To address this issue, several study efforts have been conducted, including federated learning Hoang et al. (2023) and transfer learning Hoang and Kim (2024). However, transfer learning does not always bring a positive impact on new tasks. When there is little in common between domains, knowledge transfer may fail, leading to negative transfer and reduced performance on the target task Zhuang, Qi, Duan, Xi, Zhu, Zhu, Xiong and He (2020). Furthermore, with its decentralized data sources, federated learning faces data privacy and security challenges and high communication bandwidth needs. For IDS, the main issue is communication overhead, affecting efficiency and scalability due to the high data transmission costs during training Agrawal, Sarkar, Aouedi, Yenduri, Piamrat, Alazab, Bhattacharya, Maddikunta and Gadekallu (2022). On the other hand, universal IDS can effectively detect attacks in different vehicles, regardless of their data profile or network architecture. By further standardizing universal IDS, we can ensure system interchangeability, enhance reliability and meet regulatory requirements Jeong et al. (2023). Utilizing AI technology with wavelet feature extraction and Pearson correlation, universal IDS provides innovative vehicle capabilities and serves as a practical cybersecurity solution for the evolving automotive environment. It also improves vehicle-wide adaptability by streamlining IVN monitoring and addressing changes in data profiles and network architectures. This study demonstrates the flexibility of IDS by training two models on low-frequency infusion data from two vehicles and testing them on three different vehicles to showcase their ability to adapt to various data frequencies.

This study focuses on DoS, Fuzzing, and Replay attacks that target CAN protocol in IVNs. These attacks Islam, Sahlabadi, Kim, Kim and Yim (2023b) present unique challenges to vehicular systems and represent the most significant cybersecurity threats to IVNs. The study conducts these attacks on high and low-frequency bases since anomaly injection in IVNs is frequency agnostic Islam et al. (2023b). The frequency, nature, and structure of malicious messages significantly impact IVN anomaly detection. The study aims to understand cybersecurity threats in vehicular networks better and create more effective countermeasures against them.

**Section 2** discusses various IDS, including universal IDS and conventional IDS approaches. It also explores the application of Wavelet and Pearson methods in security. **Section 3** outlines methodologies for feature extraction for deep learning purposes and explains the preprocessing of data. **Section 4** focuses on the implementation of IDS. **Section 5** evaluates the results obtained from the implemented methodologies. **Section 6** provides a comprehensive discussion of the findings and insights of the study. Finally, **Section 7** concludes the paper by summarizing the key outcomes and contributions of the study.

## 2. Related Works

The section on universal approaches examines the limitations of existing IDSs, highlighting their constraints. The feature selection section explores the adaptability challenges of conventional IDSs in practical situations. Finally, the data conversion section focuses on the concept of data generalization.

### 2.1. Universal Approached Intrusion Detection

To the best of our knowledge, only three studies Novikova, Le, Yutin, Weber and Anderson (2020); Bozdal, Samie and Jennions (2021); Islam, Oh and Yim (2023a) have been conducted on a universal approach. Firstly, Novikova et al. proposed an unsupervised anomaly detection approach for CAN bus, identifying consistent signals across nine vehicles grouped into 32 subgroups. However, practical implementation faces challenges due to the requirement for separate autoencoder models for each subgroup, especially in resource-constrained IVN Shahriar, Xiao, Moriano, Lou and Hou (2023). Accuracy measurements like false positive rates were not provided to evaluate their IDS.

Secondly, Mehmet et al. introduced WINDS, a Wavelet-based Intrusion Detection System with a specific accuracy matrix, requiring lower computational resources compared to Novikova et al. (2020). It aims to universally enhance vehicle security against specific CAN ID injections. WINDS is rule-based, employing data generation to extract features transformed into wavelet coefficients, demanding high computational power Srivastava (2022). However, it may generate false alerts with attack frequency changes and lacks cross-validation Limbasiya, Teng, Chattopadhyay and Zhou (2022).

Thirdly, Rezanur et al. utilized a combination of heatmaps from CAN ID sequences, time gaps, and hamming distances of CAN IDs to develop a universal IDS from real vehicles. They applied various CNN architectures as IDS to find the best model. However, a concern is that the study was conducted on only one vehicle with only high-frequency injection. Therefore, further research is needed to validate the ability of the IDS to be manufacturer-independent and demonstrate universal IDS capability.

In summary, Novikova et al. conducted a study on achieving universal IDS capability. They utilized diverse datasets from different vehicles, emphasizing the importance of multi-source data. In contrast, Mehmet et al. and Rezanur et al. employed Continuous Wavelet Transform (CWT) and heat-map techniques, highlighting the significance of feature selection and data conversion. These methodologies enhance data generalization and standardization, ultimately improving adaptability and facilitating analysis across various applications in the vehicle network system. More details on this topic can be found in Section 2.2.

## 2.2. Feature Selection for Intrusion Detection

This section reviews conventional IDSs methodology and input features (CAN ID sequence-based, payload sequence-based, full data utilization-based, voltage signal-based, and hybrid-based detection) from the universal IDS point of view. Numerous studies Yu et al. (2022); Xun, Deng, Liu and Zhao (2023); Mansourian, Zhang, Jaekel and Kneppers (2023); Jedh, Othmane, Ahmed and Bhargava (2021); Lo, Alqahtani, Thakur, Almadhor, Chander and Kumar (2022) have been done on IVN security, mainly on their applicability to data sources in vehicular environments.

Yu et al. introduce TCE-IDS, employing a time interval conditional entropy algorithm to analyze decimal message IDs and data blocks from the CAN network in real-time for cyberattack detection. However, it relies on predefined rules and lacks the generalizability of a universal IDS, necessitating frequent customization for vehicle or firmware updates.

Xun et al. propose an approach focusing on voltage signals from the CAN bus, using FeatureBagging combined with CNN for identifying physical layer attacks. While effective, it may be less proficient in detecting application layer attacks Wu, Li, Xie, An, Bai, Zhou and Li (2019). Implementing an IDS at the application layer offers context awareness and content inspection advantages. However, a universal IDS may require assistance with voltage features due to diverse ECU hardware configurations.

Mansourian et al. propose a payload IDS. They utilize LSTM and ConvLSTM prediction models with a Gaussian Naïve Bayes classifier, achieving high precision in distinguishing attack-free and attack data. Challenges include detecting fuzzy attacks, resource demands on limited ECUs, and intruders using packet injection to alter CAN ID while maintaining the payload.

Jedh et al. study CAN ID sequence-based IDS, utilizing Messages-Sequence Graphs and various techniques, including cosine similarity, Pearson correlation, threshold-based methods, LSTM-RNN, and Change Point Detection (CPD). While effective, the study's limited dataset raises concerns about generalizability to diverse vehicles and driving conditions. Additionally, attackers manipulating payload while maintaining CAN ID consistency pose challenges for the IDS.

Lo et al. present HyDL-IDS, a hybrid CNN-LSTM-based intrusion detection system for IVN. While effective, it has limitations, including potential bottlenecks in LSTM's information extraction and computational overhead, especially for ECUs with limited capacity. If the primary model fails, the entire system may fail. Another limitation is that the CAN ID and payload are standardized according to CAN DBC Islam et al. (2023b), requiring customization for each vehicle.

As a result, it is important to address the limitations of current IDS systems. These systems operate only with specific data sources from particular vehicles and lack the ability to generalize data.

## 2.3. Data Generalization for Intrusion Detection

Wavelet transforms enhance data generalization by capturing both time and frequency information Guo, Zhang, Lim, Lopez-Benitez, Ma and Yu (2022), enabling a multi-resolution view that isolates key features while reducing noise Srivastava (2022); Chavez and Cazelles (2019). This transform allows for efficient dimensionality reduction and temporal localization, which improves pattern recognition in non-stationary data like signals or time series. By compressing data and retaining only essential features, wavelets help machine learning models focus on relevant patterns, leading to improved generalization and reduced overfitting.

CAN systems generate large amounts of data (shown in Table 1) that can be difficult to analyze due to its changing nature. Frequency domain analysis is a helpful approach to manage this complexity and gain a better understanding of the data. The Fourier transform is a commonly used technique in frequency domain analysis, but it has the drawback of compromising frequency and time resolution, which is problematic when dealing with non-stationary data Parsons, Boonman and Obrist (2000); Feichtinger and Strohmer (2012). This compromise is a common challenge in extensive data analysis, especially with large-scale datasets. Wavelet analysis is a powerful tool for handling evolving data. Unlike the Fourier transform, wavelet analysis is effective in dealing with non-stationary data Chavez and Cazelles (2019). It uses a wavelet function to analyze data at multiple scales, allowing the identification of broad and fine-scale patterns. This approach provides valuable insights into the frequency and time characteristics of the data Zhu, Hadzima-Nyarko and Bonacci (2021), making it particularly suitable for datasets with dynamic frequency patterns. Wavelet analysis is widely used in various domains, including image analysis, telecommunications, anomaly detection, and biomedical data analysis Prasad and Iyengar (2020); Lakshmanan and Nikookar (2006); Kwak, Han and Kim (2020); James, Hou and Li (2018); Frangakis, Stoschek and Hegerl (2001); Han, Kwak and Kim (2022). The wavelet transform builds upon the short-time Fourier transforms, offering high-frequency resolution at lower frequencies and high-time resolution at higher frequencies Kehtarnavaz (2008). This characteristic overcomes the limitations of Fourier transforms, making it valuable in extensive data analysis.

On the other hand, Pearson correlation is a useful method for detecting cyberattacks as it can identify unusual network traffic patterns Jedh et al. (2021); Hoque, Kashyap and Bhattacharyya (2017); Gottwalt, Chang and Dillon (2019). It measures the linear relationship between variables and effectively detects deviations from expected behavior in network data. Analyzing correlations between various network parameters can reveal anomalies and potentially malicious activities. The resilience to outliers and ability to capture positive and negative correlations make Pearson correlation a valuable tool for identifying subtle attack patterns in complex network environments.

Table 1
Comparative Vehicle Data Analysis Based on CAN IDs, Data Generation and Message Timing Across Different Vehicles.

| Vehicles | Num. of IDs | Data Generation (Sec) | Avg. Time Gap (Sec) |
| --- | --- | --- | --- |
| Kia (LISA) | 45 | 2085 | 0.4772 |
| Kia (HCRL) | 45 | 2085 | 0.4772 |
| Sonata (HCRL) | 27 | 1943 | 0.5132 |
| Tesla (LISA) Before firmware update | 69 | 3126 | 0.3195 |
| Tesla (LISA) After firmware update | 73 | 2750 | 0.3638 |

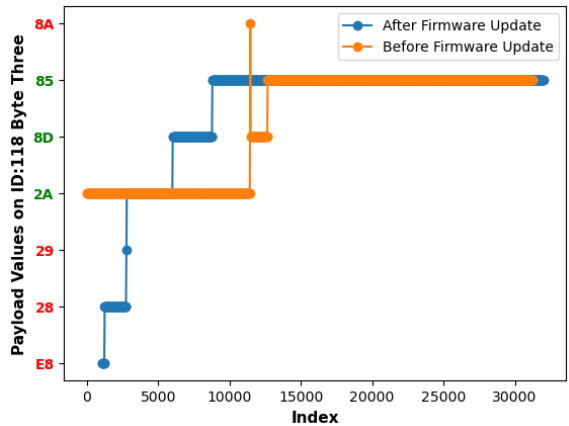

**Figure 1:** Payload Analysis on Electronic Vehicle (Tesla)

Securing the IVN is challenging and time-consuming due to the unique data profiles of each vehicle. Conventional IDS solutions require customization for each vehicle type and struggle with firmware updates and configuration changes. A universal IDS is needed to simplify security measures, eliminate the need for customization, and ensure robust protection against evolving data patterns and adversarial attacks. The solution incorporates advanced machine learning techniques and a two-stage verification process, providing a standardized and effective security solution for all types of vehicles.

## 3. Features Extraction and Data Preprocessing

### 3.1. Data Collection

Vehicle-related study requires data collection, and the On-Board Diagnostics II (OBD-II) system is a commonly used method. However, not all ECUs can be accessed via the OBD-II port, which limits the amount of data that can be collected. To overcome this limitation, researchers use the ECU Direct Approach (EDA) Koh, Kim, Kim, Oh and Yim (2022), which involves acquiring data through an internal gateway using line-tapping tools while accessing the vehicle's CAN network through an integrated central unit (ICU)Koh et al. (2022). The PEAK CAN system is the interfacing device for collecting normal driving data through the EDA method. This study employed a testbed framework incorporating real-world data to address safety concerns for attack data collection. In this controlled environment, attacks such as Fuzzing, DoS, and Replay were simulated to generate attack data for analysis and as input for deep learning models.

### 3.2. Feature Extraction

This study analyzes diverse vehicle data to detect unauthorized data injections in vehicles. Four datasets of different vehicles were examined to identify common features that can help improve vehicle security. The results show differences in internal IDs and payload data among vehicles. The CAN DBC formation causes differences between vehicles of different manufacturers and models Islam et al. (2023b). The internal data structure (See Fig. 1) may change significantly after a firmware upgrade. The study avoids relying on internal physical features to detect attacks and instead analyzes statistical features with common patterns indicating unauthorized message injections. Green values on the y-axis represent matched payload bytes, while red values represent unmatched ones after the firmware update.

Table 1 illustrates that Kia (LISA) and Kia (HCRL) exhibit similar statistical features despite sharing the same manufacturer and model. Although Sonata (HCRL) may have fewer CAN IDs, its data generation and average time gap metrics are consistent with those of other vehicles. Furthermore, the Tesla (LISA) autonomous vehicle generates data at the highest rate and with the shortest time intervals between messages, prioritizing safety and relying on numerous sensors. The amount of data generated by the Tesla vehicle varies depending on the driver's activity, resulting in a noticeable discontinuity in the statistical features shown in Fig. 2, where the data generation process of electronic vehicles fluctuates arbitrarily compared to that of mechanical vehicles.

### 3.3. Data Generalization

The aim of this study is to convert time series data obtained from a vehicle's internal systems into frequency domain. The ultimate goal is data generalization to develop a universal IDS that can be applied to a wide range of vehicles. The data is segmented into intervals of 0.01 seconds, and various metrics are calculated for each segment. These metrics include the amount of data generated and the average time gap between data packets within the segment.

$$T_{m,n} = \int_{-\infty}^{\infty} x(t)\psi_{m,n}(t)\,dt \quad (1)$$

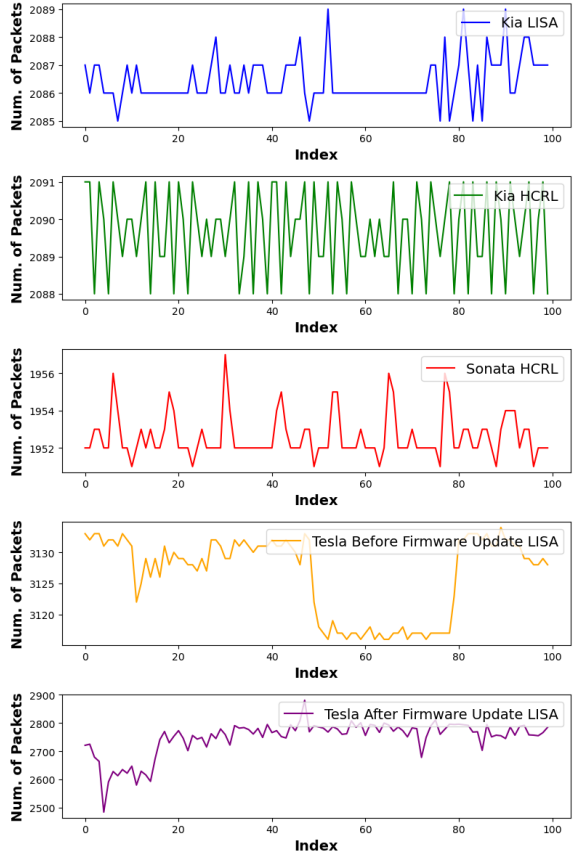

Figure 2: Data Generation Variability Among Vehicles

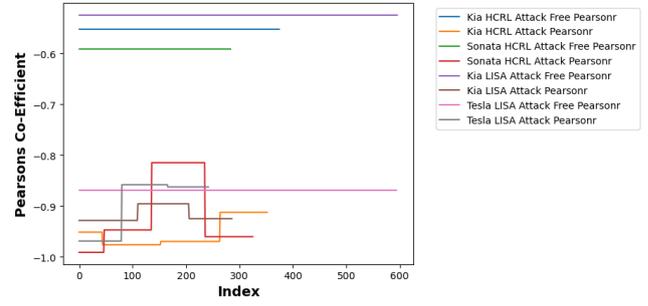

Figure 3: Pearson Coefficient Threshold

where $m$ and $n$ represent the scale and translation parameters, respectively, and $T_{m,n}$ is the wavelet coefficient that captures how closely the signal $x(t)$ matches the wavelet function at a particular scale $m$ and translation $n$. This transformation involves integrating the signal $x(t)$ over the range from negative infinity to positive infinity using the wavelet function $\psi_{m,n}(t)$. The result decomposes the signal into its frequency components, allowing for the analysis of both broad and fine-scale patterns within the signal. This process is particularly useful for analyzing time-varying data, such as IVN data, by revealing its frequency and time characteristics.

For the wavelet transform, a decomposition level of 10 and a symmetric mode were chosen, where the Daubechies wavelet with eight coefficients (db8) was used for both training and testing data. Eq. (2) and Eq. (3) represent the computation of approximation coefficients $c_{j,k}$ and detail coefficients $d_{j,k}$, respectively, at different scales and positions, where $h_n$ and $g_n$ are the scaling and wavelet coefficients of the Daubechies 8 wavelet, and $2^j$ represents the scale factor.

To apply this transform iteratively and decompose the data into multiple levels of approximation and detail coefficients, the pywt.wavedec() function was used. The symmetric parameter determined whether the input data was extended symmetrically at the boundaries before applying the transform. The level parameter specified the number of decomposition levels to be computed.

## 4. Implementation of Experimental Methods: A Comprehensive Approach

The experiment utilized the ResNet-50 model, a CNN with 50 layers and shortcut connections to enhance gradient flow, which is highly regarded for its accuracy in image classification and object recognition tasks; further details can be found in He, Zhang, Ren and Sun (2016), and the early stop method was implemented to prevent overfitting during training.

The process starts by collecting data from various CAN sources. This data is then sent through the gateway system to facilitate analysis. Timestamp alignment is used to synchronize the time intervals with the data generation events for each vehicle. For example, when generating 100 data packets, it is important to measure how long it takes for each vehicle to create these packets. By accurately aligning the

$$c_{j+1,k} = \sum_{n=-\infty}^{\infty} h_n \cdot c_{j,k+n2^j} + \sum_{n=-\infty}^{\infty} g_n \cdot c_{j,k+n2^j} \quad (2)$$

$$d_{j+1,k} = \sum_{n=-\infty}^{\infty} h_n \cdot c_{j,k+n2^j} - \sum_{n=-\infty}^{\infty} g_n \cdot c_{j,k+n2^j} \quad (3)$$

Next, a chunk of 100 values is selected for conversion into a high-resolution frequency domain representation. This process effectively converts 1 second (0.01 x 100) of statistical data into a frequency domain representation, which provides greater insight into the underlying frequency components of the CAN data. This approach allows for more detailed data analysis and can help identify patterns and trends within the CAN data.

The data conversion method used is the Discrete Wavelet Transform (DWT), and the wavelet coefficients extracted from this transformation are used as deep-learning input features due to their computational efficiency Srivastava (2022). However, challenges arise due to variations in coefficient lengths, and deep learning methods inherently require assistance in handling multi-dimensional data simultaneously. To address these issues, a "padding" technique is applied to extend shorter coefficients with zeros. This ensures uniform data dimensions for precise deep-learning analysis.

The DWT of a function $x(t)$, as described in Eq. (1), is computed with respect to the wavelet function $\psi_{m,n}(t)$,

Enhancing Automotive Security with a Hybrid Approach towards Universal Intrusion Detection System

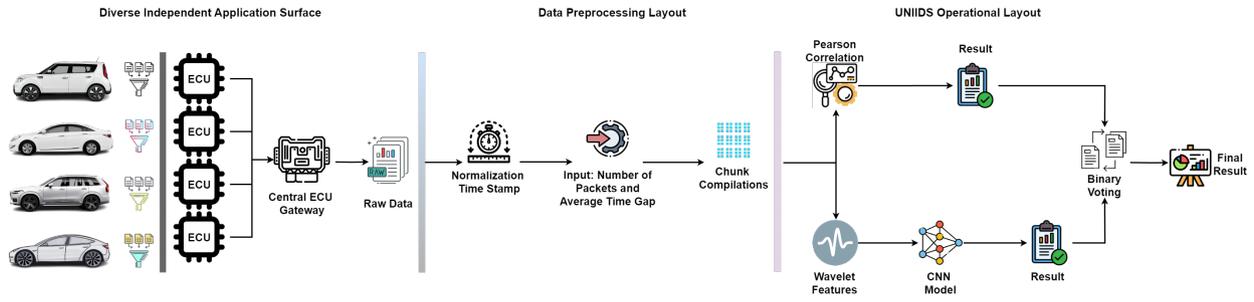

Figure 4: Abstract overview of the proposed detection system

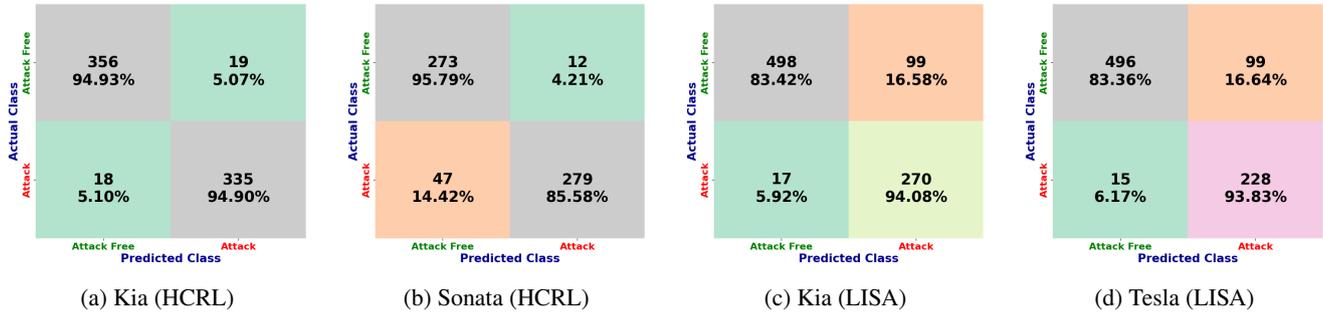

(a) Kia (HCRL)  (b) Sonata (HCRL)  (c) Kia (LISA)  (d) Tesla (LISA)

Figure 5: Confusion Matrix for ResNet-50 UIDS Training on Kia (HCRL) and Test to Different Vehicles

timestamps, the data remains consistent and reliable, especially in applications where timing is critical. It also affects the response time of IDS. Once the timestamps are synchronized, the system quantifies packet counts and calculates the average time intervals between them. These features are then subjected to wavelet conversion to enhance their ability to detect attacks from unknown sources. A CNN model based on the ResNet-50 architecture is used for binary class classification in intrusion detection. The working principle process is illustrated in Algorithm 1. Initially, Algorithm 1 uses an initial chunk size of 0.01 seconds, which may vary for each vehicle based on their data generation rates. For Sonata, the chunk size is fixed at 0.01 seconds as a reference due to its relatively lower data generation rate. For Kia and Tesla, the chunk limits are set at 0.009 seconds and 0.0065 seconds, respectively. Subsequently, a secondary chunk with a fixed length of 100, approximately 1 second of data (varied vehicle to vehicle), is applied. This data is then converted into wavelets, generating ten wavelet coefficients as deep learning features. Finally, deep learning preprocessing is performed, and the extracted features are inputted into the model for non-linear and dynamic analysis, enabling the model to detect complex patterns and relationships within the data that are crucial for accurate predictions or classifications.

$$\rho = \frac{\text{cov}(X, Y)}{\sigma_X \sigma_Y} \quad (4)$$

This IDS has been designed with a universal approach to be compatible with all types of vehicles. However, in this experiment, it was found that the accuracy of deep learning models alone was relatively low compared to the conventional IDSs standards. To improve accuracy, a secondary rule-based IDS was added. These secondary IDS utilize the same features, data generation rate, and average

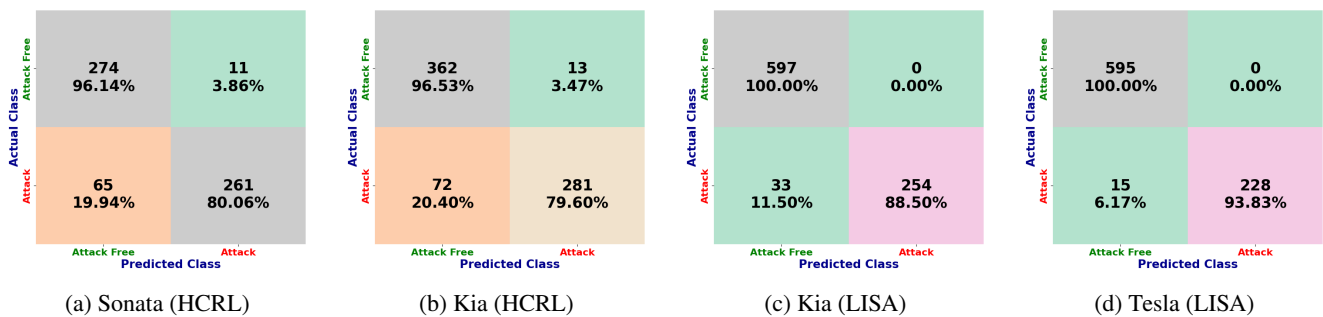

(a) Sonata (HCRL)  (b) Kia (HCRL)  (c) Kia (LISA)  (d) Tesla (LISA)

Figure 6: Confusion Matrix for ResNet-50 UIDS Training on Sonata (HCRL) and Test to Different Vehicles

**Algorithm 1:** Experimental Layout

**Input:** $data$ ▷ List of Raw Dataframes
**Output:** $Result$ ▷ Combine Result
**Step 1: Static Calculation for 0.01 Sec;**
Initialize: $S_1$, $Segment_{time}$, start, end;
**for** $i$ in length **do**
  $S_1 \leftarrow \{df_{ij} \mid start_{0.0\,ms} \leq df_{i1} \leq end_{10\,ms}, i = 1, \ldots, m\}$;
  $Segment_{time} \leftarrow S_1[\text{Num\_of\_Packets}], S_1[\text{Ave\_Time\_Gap}]$;
  start $\leftarrow$ end;
  end $\leftarrow$ end $+ 0.01$;

**Step 2: Feature Selection 0 to 100 (1 Sec);**
Initialize: $Seg_{out}$, $start_0$, $limit_{100}$, $length_N$;
**for** $i$ in range($start_0$, $length_N$, $limit_{100}$) **do**
  $Seg_{out} \leftarrow df_{input}[start_0 : limit_{100}]$;

**Step 3: Wavelet Decomposition and Scaling;**
Initialize: $wavelet_{db8}$, $mode_{sym}$, $level_{10}$;
**for** $i$ in $Seg_{out}$ **do**
  coeffs $\leftarrow$ pywt.wavedec($Seg_{out}$, $wavelet_{db8}$, $mode_{sym}$, $level_{10}$);
  $coeffs_{scaled} \leftarrow \frac{S - x_{\min}}{x_{\max} - x_{\min}} \leftarrow$ coeffs;

**Step 4: Time Series Wavelet Matrix 10 sets (1 Sec);**
Initialize: $CNN_{out}$, $X_{test}$, $y_{test}$, start, end, length;
**for** $i$ in length **do**
  $X_{test}[i] = X[start : end]$;
  $y_{test}[i] = y[end]$;
  $CNN_{out} \leftarrow CNN_{ResNet50} \leftarrow X_{test}$

**Step 5: Pearson Correlation Calculation;**
Initialize: $Pearson_{out}$, correlation, p-value;
**for** $i$ in $Segment_{time}$ **do**
  correlation, p-value = pearsonr($i$[Num_of_Packets], $i$[Ave_Time_Gap]);
  **if** correlation $\leq -0.7$ **then** $Pearson_{out} \leftarrow 1$;
  **else** $Pearson_{out} \leftarrow 0$;

**Step 6: Binary Voting;**
Initialize: $Final_{out}$;
**for** $i_1, i_2$ in $CNN_{out}, Pearson_{out}$ **do**
  **if** $i_1 == 0 \land i_2 == 0$ **then**
    $Final_{out} \leftarrow$ Attack Free;
  **else** $Final_{out} \leftarrow$ Attack;

time gap through Pearson correlation. Pearson correlation is a statistical measure, also known as Pearson's correlation coefficient, that helps determine the strength and direction of the linear relationship between two continuous variables. The coefficient ranges from -1 to 1, where 1 indicates a perfect positive correlation, -1 indicates a perfect negative correlation, and 0 denotes no linear correlation Jedh et al. (2021).

The Pearson correlation coefficient, denoted by $\rho$, measures the strength and direction of the linear relationship between two variables, X and Y. The term $cov(X, Y)$ represents the covariance, which quantifies how much X and Y change together. The standard deviations of X and Y, denoted as $\sigma_X$ and $\sigma_Y$ respectively, indicate the variability within each variable. A positive $\rho$ suggests a positive correlation, meaning that as one variable increases, the other tends to increase, and vice versa. Conversely, a negative $\rho$ indicates a negative correlation, where one variable tends to decrease as the other increases. In Eq. 4, attack-free and attack data are inputted separately to determine the threshold. Fig. 3 illustrates highly negative correlations between data generation and time gap in the attack data. When considering attack-free data for Kia and Sonata, the coefficient range is set above -0.6. However, for attack data, the coefficient range is set below -0.8. In the case of Tesla, the attack-free data crosses the threshold due to the higher data generation amount compared to mechanical vehicles, and the data generation amounts have a discontinuity in everyday driving situations portrayed in Fig. 2. This issue significantly affects the time gap sequence, resulting in a corresponding reaction in the covariance of Pearson. The combined IDS uses two independent algorithms, shown in Fig. 4, to determine whether data is an attack through a binary voting process. If both algorithms output 0, the data is considered attack-free. On the other hand, if either of the algorithms outputs a 1, the IDS classifies the data as an attack.

This Fig.4 illustrates the over all processing pipeline for UIDS. It shows multiple vehicles collecting raw data, which is then passed through preprocessing stages, likely filtering or feature extraction mechanisms. The preprocessed data is then fed into both Pearson correlation analysis and a wavelet transformation. The wavelet-transformed data is subsequently input into a ResNet-50 model. Afterward, the outputs from both models are combined using a binary voting mechanism to make a final decision, ensuring more robust and accurate results.

## 5. Result Evaluation

The study utilizes two frequency injections: low-rate periodic injections from HCRL and high-frequency injections from LISA. The study aims to evaluate the performance of a machine learning algorithm trained on two different vehicle datasets, Sonata and Kia, using low-frequency periodic injection during the training phase. The algorithm's effectiveness was tested in four scenarios: Sonata to Sonata low-frequency periodic injection, Sonata to Kia low-frequency periodic injection, Sonata to Kia high-frequency injection, and Sonata to Tesla high-frequency injection. The same approach was applied to the Kia-trained algorithm.

The performance evaluation of a deep-learning model is shown in Fig. 5 and 6, as well as Table 2. These results specifically pertain to the deep-learning model's ability to detect attacks in various vehicle data scenarios. When

**Table 2**
Performance metrics of UIDS (ResNet-50) across different vehicle datasets for low and high frequency injection attacks.

| Train Vehicle | Low Frequency Periodic Injection | | | | | | High Frequency Injection | | | | | |
|---|---|---|---|---|---|---|---|---|---|---|---|---|
| | Sonata Test (HCRL) | | | Kia Test (HCRL) | | | Kia Test (LISA) | | | Tesla Test (LISA) | | |
| | F1 Score | Acc | AUC | F1 Score | Acc | AUC | F1 Score | Acc | AUC | F1 Score | Acc | AUC |
| Kia (HCRL) | 0.90 | 0.90 | 0.91 | 0.95 | 0.95 | 0.95 | 0.87 | 0.87 | 0.89 | 0.87 | 0.87 | 0.89 |
| Sonata (HCRL) | 0.88 | 0.88 | 0.88 | 0.89 | 0.88 | 0.88 | 0.96 | 0.96 | 0.94 | 0.99 | 0.98 | 0.97 |

**Table 3**
Performance metrics of hybrid UIDS (ResNet-50 + Pearson) across different vehicle datasets for low and high frequency injection attacks.

| Train Vehicle | Low Frequency Periodic Injection | | | | | | High Frequency Injection | | |
|---|---|---|---|---|---|---|---|---|---|
| | Sonata Test (HCRL) | | | Kia Test (HCRL) | | | Kia Test (LISA) | | |
| | F1 Score | Acc | AUC | F1 Score | Acc | AUC | F1 Score | Acc | AUC |
| Kia (HCRL) | 0.98 | 0.98 | 0.98 | 0.97 | 0.97 | 0.97 | 0.90 | 0.90 | 0.92 |
| Sonata (HCRL) | 0.98 | 0.98 | 0.98 | 0.98 | 0.98 | 0.98 | 1.00 | 1.00 | 1.00 |

trained on attack-free data for Kia and Sonata, the detection accuracy was approximately 96%, as demonstrated in Fig. 5 (a) and (b), using low-frequency injection training. However, the accuracy dropped to around 95% for detecting malicious data in Kia, and 86% for Sonata. In the case of high-frequency injection, the accuracy of detecting attack-free data for Kia and Tesla was relatively low at 84%, as seen in Fig. 5 (c) and (d). Despite this, malicious data was accurately detected at approximately 94% for both vehicles. Moving on to the accuracy matrix presented in Table 2, the model trained with Kia achieved an accuracy rate of 90-95%

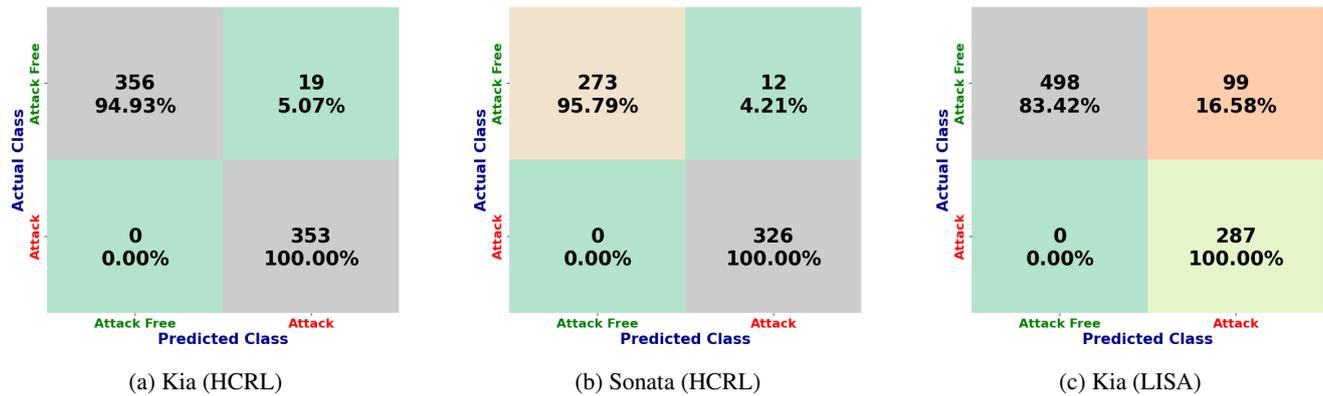

**Figure 7:** Confusion Matrix for Hybrid UIDS Training on Kia and Test to Different Vehicles

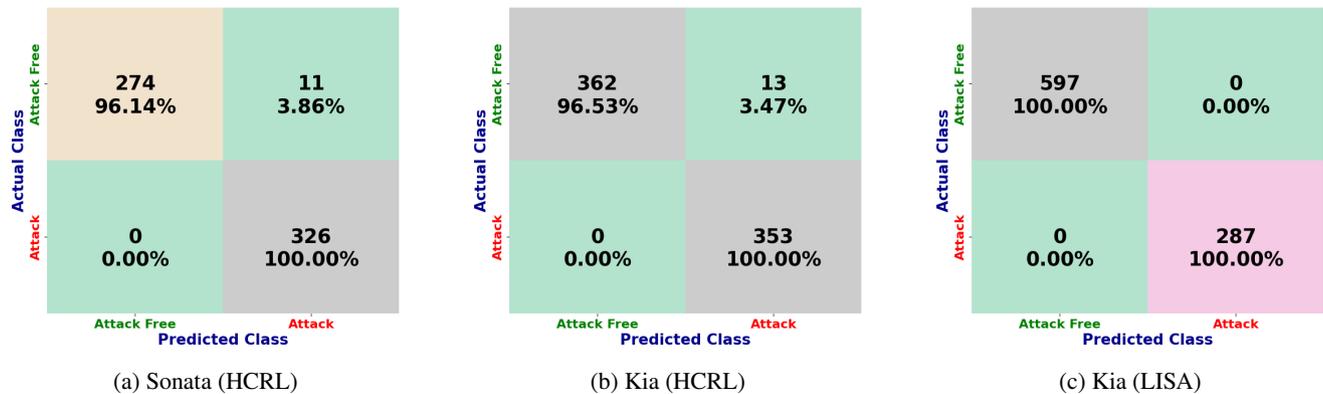

**Figure 8:** Confusion Matrix for Hybrid UIDS Training on Sonata and Test to Different Vehicles

**Table 4**
Comparative Analysis of IDS Techniques and Performance Across Various Studies.

| Approaches | Author | Data Sets | Features | Algorithm | Performance |
|---|---|---|---|---|---|
| Conventional | Lo et al. (2022) | HCRL | Full Data | CNN, LSTM | 100% |
| | Mansourian et al. (2023) | HCRL | DLC, Payload | LSTM-GNB, ConvLSTM-GNB | 100% |
| | Jedh et al. (2021) | Authors | CAN ID | Pearson, LSTM | 98.45% |
| | Yu et al. (2022) | HCRL, Authors | CAN ID, Time gap | TCE-IDS rule-based | 99% |
| | Xun et al. (2023) | Authors | Voltage Signals | SVDD | 97% |
| Universal | Bozdal et al. (2021) | Synthetic Data | Entropy | WINDS rule-based | 98% |
| | Novikova et al. (2020) | Authors, SynCAN | Signal Features | Autoencoder | Signature |
| | Islam et al. (2023a) | LISA | CAN ID, Timegap, Hamming Distance | CNN | 99% |
| Proposed | Rezanur et al. | HCRL, LISA | Entropy, Timegap | CNN | **97%-100%** |

for low-frequency periodic attacks, and 87%-89% for high-frequency injection attacks.

In comparison, Sonata trained model outperformed Kia trained model in detecting low-frequency injection attacks, as shown in Fig. 6 (a) and (b). Sonata achieved a detection accuracy of over 96% for attack-free data for both vehicles, while Kia's accuracy for attack data was only approximately 80%. However, Sonata had a 100% accuracy in detecting attack data. For high-frequency injection attacks, both Kia and Tesla had detection accuracies ranging from 95% to 98% for attack-free data, as depicted in Fig. 6 (c) and (d). Similarly, both vehicles achieved close to 100% accuracy in detecting malicious data. The accuracy rates of the deep-learning model, as presented in Table 2, were 88% for low-frequency periodic attacks and 96%-99% for high-frequency injection attacks when trained with Sonata data. These results demonstrate the effectiveness of the deep-learning model in detecting attacks in various vehicle data scenarios. In order to improve the overall accuracy of the universal IDS, which currently has relatively modest accuracy compared to conventional IDSs standards, a hybrid approach has been introduced. This approach involves combining a deep-learning model with an additional IDS.

A hybrid IDS was developed by combining a rule-based (Pearson correlation coefficients) IDS with a deep-learning model. The performance evaluation of the hybrid IDS is presented in Fig. 7, Fig. 8, and Table 3. For low-frequency injection training with Kia, as shown in Fig. 7 (a) and (b), the detection accuracy for Kia and Sonata attack-free data is approximately 96%. In comparison, malicious data was detected with 100% accuracy. In the case of high-frequency injection, as depicted in Fig. 7 (c), Kia attack-free data were accurately detected at around 84%, with malicious data accurately detected at 100%. Referring to the accuracy matrix presented in Table 3, the model trained with Kia achieved 98% and 97% accuracy for low-frequency periodic and 91% high-frequency injection, respectively. Compared to Sonata, Kia's accuracy performance showed a decrease of less than 1%.

In the case of low-frequency injection training with Sonata, as shown in Fig. 8 (a) and (b), the detection accuracy for Sonata and Kia attack-free data is approximately 97%. On the other hand, malicious data was detected with a 100% accuracy rate. For high-frequency injection, as depicted in Fig. 8 (c), Kia attack-free data were detected accurately at 100%, with malicious data also accurately detected at 100%. Referring to the accuracy matrix presented in Table 3, the model trained with Sonata achieved an accuracy of 98% for low-frequency periodic injection and 100% for high-frequency injection. When comparing our results with other study, the overall accuracy meets the benchmark compared to conventional IDS and universal IDS shown in Table 4.

## 6. Discussion

Based on the evaluation of the experimental results, it has been observed that injecting data at low frequencies can lead to mispredictions. This is because the injected data is generated periodically and with a relatively small volume, causing the system to behave like a typical data flow. As a result, misclassification can occur. On the other hand, high-frequency injections on Kia-trained model to mistakenly predict attack-free data as an attack. This misidentification only happens in the case of replay attacks, which have similar characteristics to attack-free data. Two advanced deep-learning models have been implemented in intrusion detection systems: one from Kia and the other from Sonata. Remarkably, Kia's model is highly effective in detecting low-frequency injection attacks, while Sonata's model excels at detecting high-frequency injection attacks for mechanical vehicles. Interestingly, both hybrid model from Kia and Sonata perform equally well for mechanical vehicles. For optimal deployment of IDS in real-world scenarios, it is advisable to use a models with varying data injection frequencies.

During testing on Tesla, the hybrid model shows limited performance due to a discontinuity in data generation (see Fig. 2). The differing data generation processes between mechanical and electronic vehicles influence the time gap between packets, significantly impacting the Pearson correlation coefficient (see Fig. 3). While Pearson correlation effectively identifies linear relationships and anomalies in network traffic, it struggles with the complex, non-linear patterns typical of CAN bus data, where data generation varies with driving conditions and driver behavior, resulting in shifting temporal and spatial correlations. ResNet-50 mitigates this limitation by using deep learning to capture non-linear features, accommodating the dynamic nature of vehicle datasets and improving intrusion detection accuracy. In mechanical vehicles, omitting Pearson correlation reduces prediction accuracy by approximately 10%, underscoring a trade-off between accuracy and broader scalability. By combining Pearson's statistical insights with ResNet-50's capacity to adapt to evolving data patterns, the system enhances detection capabilities for mechanical vehicles, capturing both straightforward and intricate patterns across varied driving scenarios. Future studies should focus on developing UIDS electric vehicle-specific attack detection methods that optimize accuracy and scalability, strengthening detection robustness across a broader range of electric vehicle applications.

## 7. Conclusion

In conclusion, the various network architectures and data generation processes used by different car manufacturers pose challenges for conventional IDS systems. A universal IDS system offers a standardized solution to address these differences and enhance cybersecurity in the ever-changing automotive industry. The experiments thoroughly investigated the impact of data injection frequency on IDS performance, particularly in relation to variations in manufacturer IVN. Tesla's electronic vehicle model experienced performance issues due to disparities in data generation, while advanced hybrid models from Kia and Sonata effectively detected low- and high-frequency attacks for mechanical vehicles, respectively. To optimize IDS deployment in real-world scenarios, using a combination of models with different data injection capabilities is recommended. Further study is needed to understand the electronic vehicle data generation process and how attacks affect IVN performance to improve data generalizability and optimize accuracy. Universal IDS emerges as a future-proof solution, ensuring a secure and consistent cybersecurity solution for advanced vehicles and prioritizing safety for all.

## CRediT authorship contribution statement

**Md Rezanur Islam:** Methodology, Analysis, Model Deployment. **Mahdi Sahlabadi:** Original draft preparation. **Keunkyoung Kim:** Original draft preparation. **Kangbin Yim:** Conceptualization of this study, review, editing and supervision.

## Data available on request from the authors

The data that support the findings of this study are available from the corresponding author upon reasonable request.

## Acknowledgment

This research was supported by the MSIT(Ministry of Science and ICT), Korea, under the Convergence security core talent training business support program (IITP-20242710008611) supervised by the IITP (Institute for Information Communications Technology Planning Evaluation) and Soonchunhyang University Research Fund.

**Md Rezanur Islam** received his Bachelor of Science in Electrical and Electronic Engineering from The University of Asia Pacific in Bangladesh in 2016. He later pursued a Master of Science in Mobility Convergence from Soonchunhyang University in South Korea in 2023. Currently, he is a Ph.D. student in Software Convergence at Soonchunhyang University in South Korea. His research focuses on deep learning, anomaly detection, malware detection, computer vision, and NLP, with a focus on driver state recognition, reflecting his commitment to investigating advanced solutions and leveraging state-of-the-art technologies in these domains.

**Mahdi Sahlabadi**, an IEEE Senior Member, holds a Ph.D. in Industrial Computing from the National University of Malaysia. His academic journey includes research positions at the Japan Advanced Institute of Science and Technology (JAIST), Singapore Management University (SMU), Sharif University of Tehran (SUT), University Kebangsaan Malaysia (UKM), and Soonchunhyang University (SCH), South Korea. His areas of research interest are process mining, software architecture, cybersecurity, and quality assurance.

**Keunkyoung Kim** received her B.S. and M.S. degrees from the Department of Electronics Engineering, Ajou University, Suwon, South Korea, in 1999 and 2002, respectively. She is currently pursuing her Ph.D. degree at Soonchunhyang University in the Department of Software Convergence Engineering. Her research interests include big data analysis and deep learning technology for malicious packet filtering and misbehavior detection.

**Kangbin Yim** is a professor in the Department of Information Security Engineering at Soonchunhyang University, where he has been since 2003. He received his B.S., M.S., and Ph.D. degrees in Electronics Engineering from Ajou University, South Korea, in 1992, 1994, and 2001, respectively. For over 20 years, his primary research has focused on vulnerability identification, threat analysis, and proof-of-concept (PoC) development for both software and hardware. He is also passionate about designing and implementing hardware and software frameworks for system evaluations and commercial services. His recent work has centered on HILS-based dynamic analysis for distributed embedded software, leading a research team of over 30 members in his lab, LISA. Currently, the lab's top priorities include deep-learning-driven analysis of heterogeneous field data with a particular focus on automotive vehicles, industrial control systems, and mobile baseband.